# Anisotropic electron-phonon interactions in 2D lead-halide perovskites

Jaco J. Geuchies[1*], Johan Klarbring[2,3], Lucia Di Virgillio[1], Shuai Fu[1], Sheng Qu[1], Guangyu Liu[4], Hai Wang[1], Jarvist M. Frost[4], Aron Walsh[2], Mischa Bonn[1*] and Heejae Kim[1,5*]

1. Max Planck Institute for Polymer Research, 55128 Mainz, Germany.

2. Department of Materials, Imperial College London, London SW7 2AZ, United Kingdom.

3. Department of Physics, Chemistry and Biology (IFM), Linköping University, SE-581 83, Linköping, Sweden

4. Department of Physics, Imperial College London, London SW7 2AZ, United Kingdom.

5. Department of Physics, Pohang University of Science and Technology, 37673, Pohang, Korea.

*Corresponding authors

KEYWORDS (Word Style "BG_Keywords"). Low-dimensional perovskites, electron-phonon coupling, polarons, ultrafast THz spectroscopy.

**ABSTRACT**

Two-dimensional (2D) hybrid organic-inorganic metal halide perovskites offer enhanced stability for perovskite-based applications. Their crystal structure's soft and ionic nature gives rise to strong interaction between charge carriers and ionic rearrangements. Here, we investigate the interaction of photo-generated electrons and ionic polarizations in single-crystal 2D perovskite butylammonium lead iodide (BAPI), varying the inorganic lamellae thickness in the 2D single crystals. We determine the directionality of the transition dipole moments (TDMs) of the relevant phonon modes (in the 0.3 – 3 THz range) by the angle- and polarization-dependent THz transmission measurements. We find a clear anisotropy of the in-plane photoconductivity, with a ~10% reduction along the axis parallel with the transition dipole moment of the most strongly coupled phonon. Detailed calculations, based on Feynman polaron theory, indicate that the anisotropy originates from directional electron-phonon interactions.

# INTRODUCTION

Hybrid lead-halide perovskites are one of the first ionic semiconductors where the diffusion of the photoexcited charge carriers exceeds 1 micron[1]. Several of the advantageous properties of hybrid perovskites, including reduced carrier-impurity interactions[2], long charge carrier lifetimes[3], and high defect tolerance[4] have been associated with the material's distinctive electron-phonon interactions. These interactions also set a limit for the maximum carrier density and mobility in perovskites[5]. Despite outstanding characteristics, the instability of the halide perovskites in an ambient environment has hindered its applicability. One developmental direction is reducing the connectivity of the octahedral networks in the perovskite structure to two dimensions (2D). Since 2D perovskites exhibit enhanced stability and enable improved performance of 2D/3D heterostructures[6–9], it would be ideal to systematically investigate potentially interesting optoelectronic properties arising from the confinement.

The perovskite structure can be transformed from 3D into 2D by replacing some of the smaller A-site cations (e.g. methylammonium or formamidinium) with organic molecules with longer C-chains[10], such as butylammonium (BA)[11], phenylethylammonium and the chiral methylbenzylammonium[12,13]. These longer spacer molecules push the lattice apart into two-dimensional metal-halide lamellae with tunable thicknesses, which alters the energetic landscape of photoexcited carriers due to quantum- and dielectric confinement[14]. The vibrational properties of low-dimensional perovskites also differ from their 3D counterparts. Due to a large impedance mismatch between the inorganic and organic layers in 2D perovskites, acoustic phonon propagation, i.e., heat transport, is two times slower compared to 3D perovskites[15]. Butylammonium lead iodide (BAPI) perovskites have been studied extensively[16,17], but especially when synthesized through a synthesis protocol involving spin coating[11], it is hard to directly obtain large-area crystals which are phase pure in terms of the thickness of the inorganic structure *n*. In small flakes, room-temperature Raman microspectroscopy measurements in literature have revealed in-plane anisotropy of the Raman-active modes[18], which were shown to be strongly heterogeneously broadened upon lowering the temperature. The broadening of the individual modes as a function of temperature is heavily related to the strong anharmonicity of the halide perovskite phonon modes, in particular via coupling to octahedral tilting[19]. Due to the small size of the perovskite crystal flakes, these experiments from literature were done microscopically, hampering direct access to optically active (absorptive) modes, and their anisotropy, in the THz frequency range.

The vibrational properties of 2D perovskites are expected to play the same crucial role in determining electronic properties as in 3D halide perovskites[20–22]. In particular, thermally accessible vibrational modes (up to 25 meV (6 THz) at room temperature) of perovskites originate mostly from displacements of the inorganic framework[23]. Previous studies have shown that these low-frequency phonon modes couple strongly to electronic states around the band extrema of the 3D perovskite[24], which can be rationalized by the fact that these electronic bands are made from a basis of lead- and halide atomic orbitals. Specifically, in 3D methylammonium lead iodide (MAPI), the displacement of the structure along the 1 THz phonon coordinate was revealed as the mode that dominates electron-phonon coupling and has been used to explain the temperature dependence of the bandgap of MAPI[24–26]. Although the exact structure of the polaron in metal halide perovskites is still under debate[27], it is clear that the coupling between vibrations and charge carriers in these materials impacts their static optoelectronic properties and dynamic phenomena, e.g., carrier cooling[28,29].

Anisotropic electron mobilities have been observed in, and predicted for, various semiconductor and metallic materials, such as $TiO_2$[30,31], phosphorus carbide[32], 2D niobium selenide[33], materials with tilted Dirac cones (Borophene)[34] and the organic semiconductor tetracene[35]. In tetracene, the anisotropy of the electron mobility is directly correlated with the vibrational properties of the material. In 2D perovskites, depending on the exciton-binding energy, there is a subtle balance between the formation of excitons and free carriers[36]. Perhaps counter-intuitively, the diffusivity of excitons in-between the inorganic layers is efficient due to dipolar energy-transfer mechanism[37,38]. While five times slower than intra-layer transport, this is less anisotropic than charge-carrier transport, which proceeds via exponentially decaying electron wavefunction overlap. While previous studies on transport mechanisms in 2D perovskites have focused on transport anisotropy between the inorganic layers[37,38], little is known about anisotropy in carrier diffusion inside the inorganic layers[39] and its direct relation with electron-phonon interactions.

It is therefore important to investigate the anisotropic nature of electron transport inside the inorganic lamella, whether or not it is inherited from coupling with directional phonon TDMs, as well as the effect of confinement on both electronic and vibrational properties. In this work, we systematically study the anisotropy of electron transport within the inorganic lamella, the potential contribution from coupling with specific phonons, as well as the effect of confinement on both electronic and vibrational properties. To this end, we synthesize large-area butylammonium lead iodide single crystals, with precise control over the inorganic layer thickness, $n$, and investigate the electron conduction mechanisms and optically active vibrational modes in the 0-3 THz spectral range with polarized THz time-domain spectroscopy (TDS) at a fixed level of confinement. We identify the crystallographic direction of the TDMs of all optically active phonon modes, and observe the time/frequency-resolved photoconductivity along, and perpendicular to, the direction of the 1 THz phonon TDM (i.e. the [102] and [201] directions). We observe a clear anisotropy in the photoconductivity, where the mobility is ~5-10% larger/smaller in perpendicular/parallel direction. Combining density functional theory (DFT) simulations and Feynman polaron theory, we unveil that the apparent anisotropy in photoconductivity originates from anisotropic electron-vibration interactions with the 1 THz phonon mode.

The results here shed light on anisotropy in optically active vibrational modes and carrier-transport properties in single-crystalline 2D perovskite materials and open the door to designing devices relying on enhanced electron mobilities along specific crystallographic directions.

RESULTS AND DISCUSSION

**Steady-state structural and optical characterization.**

We used a synthesis protocol adapted from literature[40] (see methods section SI1 in the SI) to grow BAPI single crystals with a large lateral area (> 1cm$^2$) and high $n$ purity, sufficiently thin (~10-200 m) to transmit photons in the THz frequency range. Large-area single crystals are a prerequisite for probing anisotropy through THz spectroscopy due to the diffraction-limited size of our THz pulse (~1 mm diameter). Figure 1(a) shows photographs of representative crystals at the liquid-air interface of the precursor solution. BAPI crystals consist of layered, primarily inorganic sheets, electronically decoupled by BA. By changing the ratio of BA and MA ions, we control and vary the thickness of the inorganic structure between $n$ = 1 - 4. The changing bandgap in the visible spectral range going from $n$ = 1 to $n$ = 4, due to the weakening of both quantum- and dielectric confinement, can be observed by eye. The bottom part of the figure shows the extended unit cells of the BAPI crystals, where we have used the convention to denote directions <h 0 l> in the inorganic planes of the structure. The high phase purity of the BAPI crystals is shown both by the powder X-ray diffraction (pXRD) measurements in Figure 1(b) as well as the reflectivity spectra, shown in Figure 1(c). From the pXRD, we determine the inter-lamellar spacing, $d_{010}$, to increase from 13.77 ± 0.04 Å for $n$ = 1 to 32.17 ± 0.05 Å for $n$ = 4, an increase of 6.1 ± 0.2 Å/$n$, i.e., precisely the size of one methylammonium lead iodide octahedron (see SI section SI2).

The reflectivity spectra in Figure 1(c) for all different layer thicknesses show two features. A transition at higher energy, which corresponds to the excitonic absorption line, and a sub-bandgap feature at longer wavelengths, both of which are indicated by vertical dashed lines in the spectra. The sub-bandgap absorption line, which can also be observed in emission, has been ascribed to various origins, from states localized at the edges of crystals to magnetic dipole emission[41–44]. A full evaluation and assignment of the reflectivity curves is beyond the scope of this work.

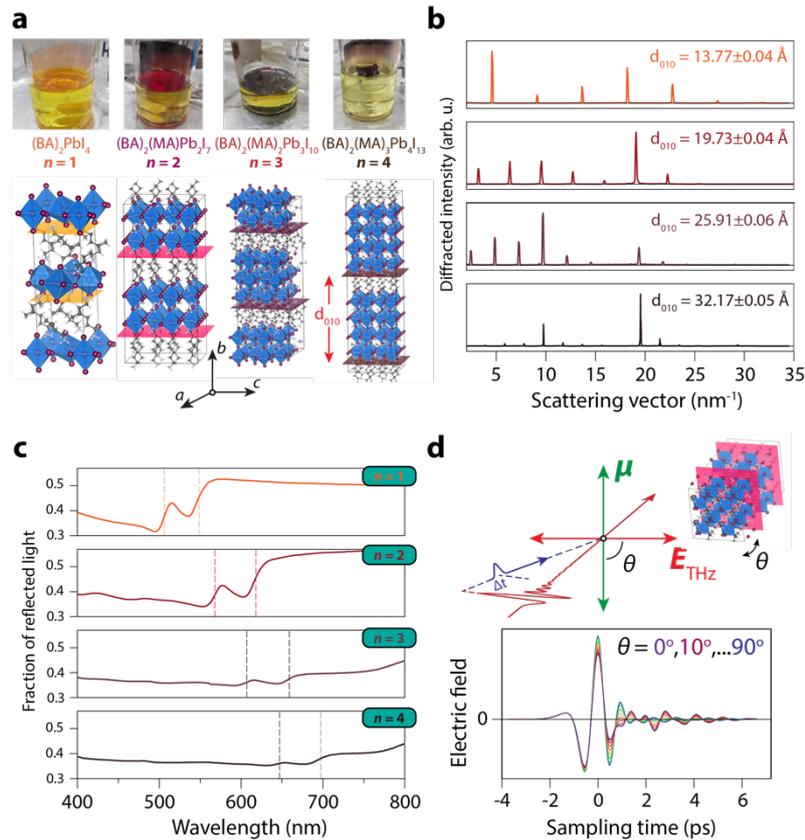

**Figure 1: Structure and linear optical properties of two-dimensional single-crystalline BAPI and schematics of the polarization-resolved ultrafast THz spectroscopy. (a)** Top row - photographs of single-crystalline $(BA)_2PbI_4$ ($n$ = 1), $(BA)_2(MA)Pb_2I_7$ ($n$ = 2), $(BA)_2(MA)_2Pb_3I_{10}$ ($n$ = 3) and $(BA)_2(MA)_3Pb_4I_{13}$ ($n$ = 4). Bottom row – extended unit cells of 2D BAPI. **(b)** Powder XRD of single-crystalline BAPI sheets with different n, revealing

the lamellar structure and the high '*n*-purity' of the crystals. **(c)** Reflectance spectra for the 2D BAPI crystals with different *n*, showing the excitonic and sub-gap resonances. **(d)** Schematic of the polarization-resolved linear THz time-domain spectroscopy (TDS) and optical-pump/THz-probe (OPTP) experiments. As we rotate the crystal, the relative orientation of the THz electric field vector, $\mathbf{E}_{THz}$, with respect to transition dipole moments of vibrational-modes, changes. When $\mathbf{E}_{THz} \parallel$ , phonons can absorb the THz field, when perpendicular, they cannot. The bottom panel shows a transmitted THz pulse through an *n* = 1 BAPI single crystal, as a function of the azimuthal angle, θ, of the crystal. In OPTP experiments, the THz field will accelerate photoexcited carriers along its electric field polarization, probing the mobility along that axis.

Figure 1(d) shows a schematic of the polarized THz (pump-probe) spectroscopic experiments. We measure the transmission of a linearly polarized THz pulse (with frequencies between 0.3 – 3 THz) through our sample, where the beam impinges on the BAPI crystals along the surface normal of the inorganic lamellae. By rotating our sample, we effectively align the transition dipole moments (TDMs, ) in the plane of the crystal w.r.t. the electric field vector of our THz photons (with θ being the azimuthal angle between $\mathbf{E}_{THz}$ and ). When $\mathbf{E}_{THz} \parallel$ (θ = 0º), the field couples to this mode and gets attenuated; when they are perpendicular (θ = 0º), they cannot interact and the field is simply transmitted. An example of transmitted THz fields through an *n* = 1 BAPI crystal for different θ is shown in Figure 1(d). We also performed optical-pump/THz-probe (OPTP) measurements, where we photoexcite our samples at 400 nm, and probe the optical conductivity via the change in the transmitted THz electric field, $-\Delta E/E_0$, with ΔE the difference between the photoexcited and unexcited BAPI crystal, and $E_0$ the steady-state transmittance of the THz pulse. Here, the charge carriers are accelerated by the THz field in a specific crystallographic direction in the crystals, given by θ. As explained further throughout the text, we measured the OPTP transients parallel and perpendicular to the 1 THz TDM, as the modes at this frequency have been shown to dominate electron-phonon interaction[24–26,45].

**Linear THz absorption – TDS measurements of low-frequency vibrations in BAPI.**

We first discuss linear TDS measurements through the BAPI crystals. Figure 2(a-d) show the absorption spectra in the 0.3 – 3 THz frequency range. For *n* = 1 BAPI, shown in Figure 2(a), there are four distinct modes. Compared to the THz absorption spectrum of bulk MAPI, which has two modes at 0.9 THz (octahedral rocking mode) and 2 THz (Pb-I-Pb stretch)[23,46], it seems the modes are split. This can be rationalized due to the lower symmetry of the *n* = 1 BAPI crystal; the in-plane and out-of-plane vibrations in BAPI sample different potential energy surfaces compared to bulk MAPI. Indeed, below the tetragonal-to-orthorhombic phase transition temperature, the two vibrational modes of bulk MAPI split into four, a result of the reduced crystal symmetry, lifting the degeneracy of the modes[47]. Furthermore, the absorption of the modes 0.9, 1.2 and 2.5 THz can be switched on and off by rotation over $\theta$, where the former two and the latter are out of phase (i.e., they have perpendicular TDMs). The mode at 2 THz shows no dependence on $\theta$, which can be the case if there is a high apparent degeneracy. Indeed, according to the theory presented below, this peak in the absorption spectrum comprises many different eigenmodes. As these eigenmodes all have different directionalities of their TDMs, this mode can be seen as quasi-degenerate.

When increasing *n* in Figures 2(b-d), the peak splitting we observed in *n* = 1 is reduced. For *n* = 2, we observe that the mode at 1.1 THz has a $\theta$-dependent region at the lower frequency side of the mode, and a $\theta$-independent part at the high frequency side of the mode. The 'on-off ratio' of this 1 THz mode is strongly reduced compared to *n* = 1, and decreases further for *n* = 3 and 4. For all *n*, the intense mode around 2 THz shows no $\theta$-dependence. Note that for *n* = 4, we could not synthesize sufficiently thin single crystals to achieve reasonable transmission of THz photons with frequencies above 1.6 THz.

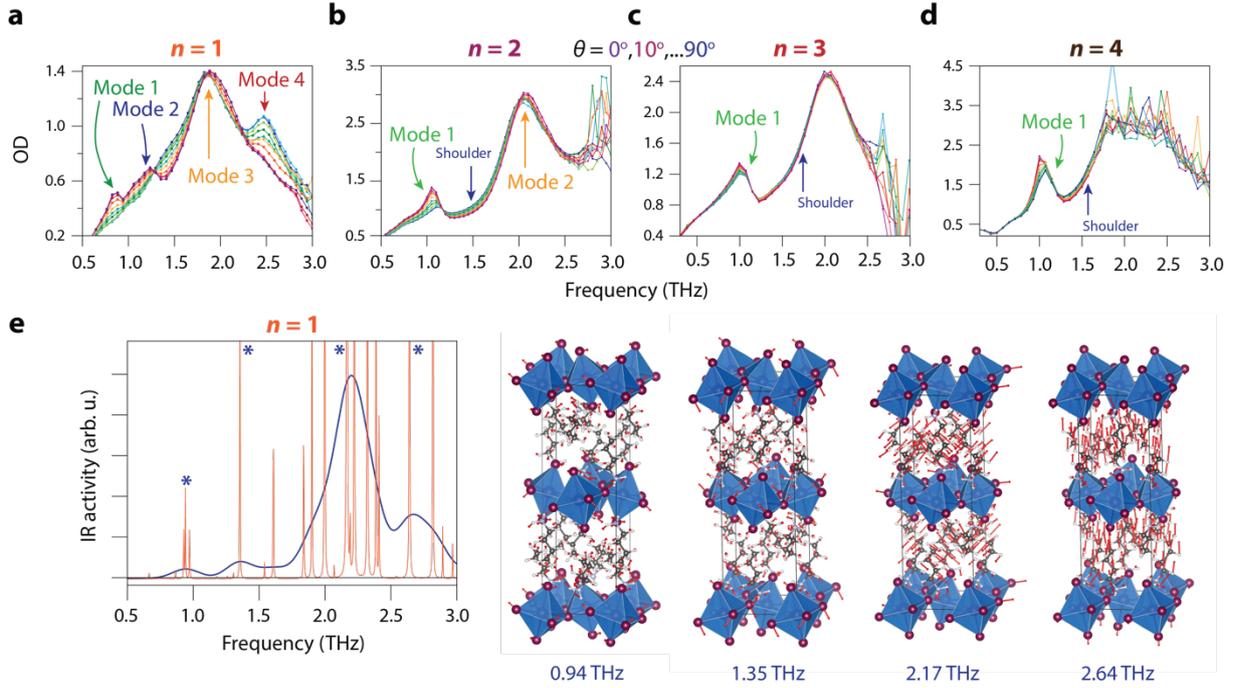

**Figure 2: THz transmission measurements through the single-crystalline BAPI crystals as a function of azimuthal angle and simulated eigenmodes for n = 1 BAPI.** (a) Extinction spectra for $n = 1$ BAPI as a function of $\theta$. Four different modes are identified. (b) Extinction spectra for $n = 2$ BAPI as a function of $\theta$. (c) Extinction spectra for $n = 3$ BAPI as a function of $\theta$. (d) Extinction spectra for $n = 4$ BAPI as a function of $\theta$. Notice how, with increasing $n$, the $\theta$-dependence and 'on/off ratio' of the modes decreases. (e) Simulated eigenmodes for $n = 1$ and the corresponding absorption spectrum. The atomic displacement vectors of the four modes with the highest oscillator strength, indicated with a (*) in the spectrum, are displayed on the right.

To understand the drastic change in the angle-dependent behavior of the vibrational spectra observed when changing the layer thickness microscopically, we computed the THz absorption spectra for $n = 1$ from DFT-based harmonic phonon theory (computational details[48–57] can be found in the SI). We separately compute spectra for the high-temperature (HT) and low-temperature (LT) phases ($T_c \approx 274K$) of n=1 BAPI based on resolved crystal structures from Ref. [19][19], (see SI fig S10). Surprisingly, the calculated spectrum of the LT phase matches the room temperature, i.e. above $T_c$, experimental data well. We believe this is due to the proposed disordered nature of the HT phase[19], i.e., that the HT phase locally resembles the LT phase, and we thus choose to use the LT structural model in the following analysis. Figure 2(e) shows the calculated IR spectrum for the LT phase convoluted with a 0.1 THz Gaussian broadening (blue line) and when a small broadening is applied to each phonon mode (orange line). The overall shape of the broadened spectrum, barring a slight blue shift, agrees well with our experimental measurements. In particular, there are four clear peaks with obvious correspondence to the four peaks of the measured extinction spectrum. It thus becomes clear that each of these peaks, in fact, contains various numbers of eigenmodes that all have significant TDMs: the peaks are heterogeneously broadened. The eigendisplacements of the phonon modes with the strongest TDMs for each peak, indicated with an asterisk in the calculated spectrum, are shown on the right-hand side of Figure 2(e). These displacements are made up of various distortions of the $PbI_6$ octahedra, but also significant translational motion of the butylammonium cations, especially for the two higher frequency modes.

Since we perform phase- and amplitude-resolved THz transmission measurements, we can directly obtain the complex refractive indices of the samples without using the Tinkham (thin-film) approximation[58]. As outlined in the SI (section SI4-5) we numerically fit the transmission functions based on Fresnel equations to those obtained from the experimental data, with the samples' refractive indices as free parameters. The results are shown in Figure S8. Through this, we can separate the absorptive component, the imaginary part of the refractive index, from the dispersive component, the real part of the refractive index. Similar to three-dimensional perovskite (MAPI)[47], the refractive indices are relatively large compared to standard semiconductors such as Si, due to the ionic and soft nature of the perovskite structure.

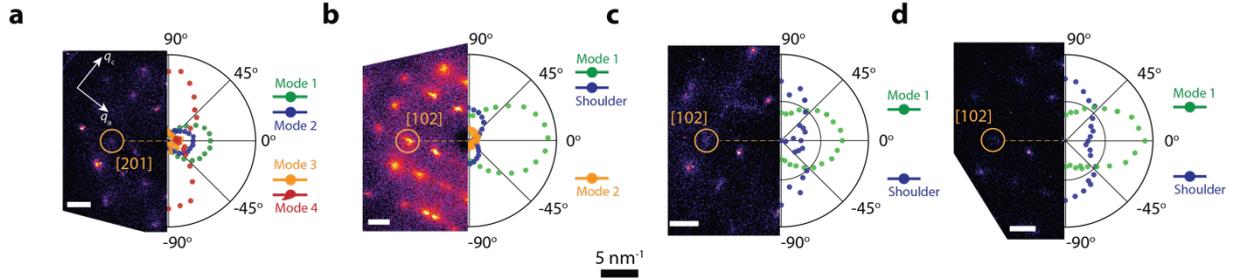

**Figure 3: transition dipole moment vectors of the vibrational modes in 2D BAPI.** (a) Single-crystal diffraction pattern of the $n = 1$ BAPI crystal (left) and the intensity of the vibrational modes (right). The transition dipole moment (TDM) of the two modes around 1 THz lies in the [201] crystallographic direction, the TDM of the 2.5 THz mode is orthogonal to the former and lies in the [-102] direction. (b) Single-crystal diffraction pattern of the $n = 2$ BAPI crystal (left) and the intensity of the vibrational modes (right). The TDM of the vibrational mode around 1 THz lies in the [102] crystallographic direction. (c) Single-crystal diffraction pattern of the $n = 3$ BAPI crystal (left) and the intensity of the vibrational modes (right). The TDM of the vibrational mode around 1 THz lies in the [102] crystallographic direction. (d) Single-crystal diffraction pattern of the $n = 4$ BAPI crystal (left) and the intensity of the vibrational modes (right). The TDM of the vibrational mode around 1 THz lies in the [102] crystallographic direction. For all different $n$, the intense mode around 2 THz shows no angle dependence.

Figure 3(a-d) shows the angle-dependence of the observable modes extracted from the data in Figure 2, correlated to the crystallographic orientation of the crystals, which we measured by performing transmission XRD on the same crystal we performed the THz experiments on. For all the thicknesses, the TDM of the 1 THz modes lies in the <2 0 1> and <1 0 2> family of directions, which differ only slightly due to variations in the unit-vector lengths in the *a* and *c* direction of the crystals. The mode at 2.5 THz is perpendicular to this direction in $n = 1$ and lies in the <-1 0 2> direction.

We have also calculated an angle-resolved IR-spectrum from our phonon simulations as shown in the SI (section SI6). We note that, since we employ the harmonic phonon approximation, the symmetry of the structure implies that the TDMs of all modes lie within the <100>, <010> or <001> family of directions. This does not match our experimental observation, and it is thus likely that higher-order effects, e.g. anharmonic phonon-phonon coupling, result in overall TDM moments in the off-diagonal <1 0 2> directions.

**Polarization resolved optical pump/THz probe measurements of the photoconductivity.**

Now that we have determined the directions of phonon TDMs of the inorganic substructures of the BAPI crystals, we compare the complex photoconductivity between directions parallel and perpendicular to the directions of the 1 THz TDMs. Figure 4(a-d) shows the real and imaginary parts of the OPTP transients along both directions after photoexcitation at 400 nm. Noticeably, the transient amplitudes exhibit anisotropy persistently for all the measured thicknesses ($n$ = 1-4) over the experimental time window (~20 ps). We note that all the OPTP measurements are performed in the low-fluence, i.e. linear, regime, to exclude the presence of higher-order recombination processes originating from carrier-carrier interactions, and to prevent photodegradation of the crystals. Furthermore, for each $n$, we kept the same excitation fluence (within 1%) for measurements parallel and perpendicular to the 1 THz TDM.

To dissect the overall carrier dynamics and conduction mechanism as well as their anisotropies, we first consider the decaying components for each thickness, $n$. For $n$ = 1-3, shown in Figure 4(a-c), the OPTP signal shows a fast decay over the first couple of picoseconds, followed by slower decay at later times. To elucidate the origin of this fast decay, we compare resonant excitation of the crystals to excitation at 400 nm, i.e. above the bandgap energy (see SI Figure S13). Upon resonant excitation, there is only an ingrowth visible, limited by the instrument response function, followed by a slowly decaying component. The OPTP traces obtained using 400 nm excitation decay over the first 1-2 ps to the same level (normalized by the absorbed photon fluence) as for the OPTP traces using resonant excitation, after which both photoconductivity traces are identical. Since the photoexcited densities are in the linear regime, the initial decay upon 400 nm excitation comes from highly mobile hot carriers, which cool within a few ps to less mobile states at the band edges, an effect previously observed for hot holes in the $Cs_2AgBiBr_6$ double perovskite[59]. A similar effect was also observed by Hutter *et al.* for methylammonium lead iodide, but ascribed to a direct-indirect character of the bandgap[60]. The comparison of resonant- and above-bandgap excitation experiments indicates that hot carriers have a higher mobility than cold carriers in BAPI, a comprehensive study of which is beyond the scope of this work. As the layer thickness increases, we expect the decay of the photoconductivity to become second-order-like, as in MAPI, where the photoconductivity does not decay over the first nanoseconds[61,62].

Next, we examined the in-plane photoconduction mechanism by recording photoconductivity spectra at a pump-probe delay time of 10 ps. We recorded photoconductivity spectra, where we measured the change in the transmitted THz pulse in the time domain and Fourier transformed it into a conductivity spectrum. Since the BAPI crystals are relatively thick (tens of micrometers), we numerically retrieved the refractive index from an analytical model of the transmission of the THz pulse through the sample[63–65]. After obtaining the steady-state and excited sample's refractive index, we converted these into dielectric functions and, from the difference, we obtain the complex photoconductivity. The full analysis and method details are outlined in the SI (section SI5).

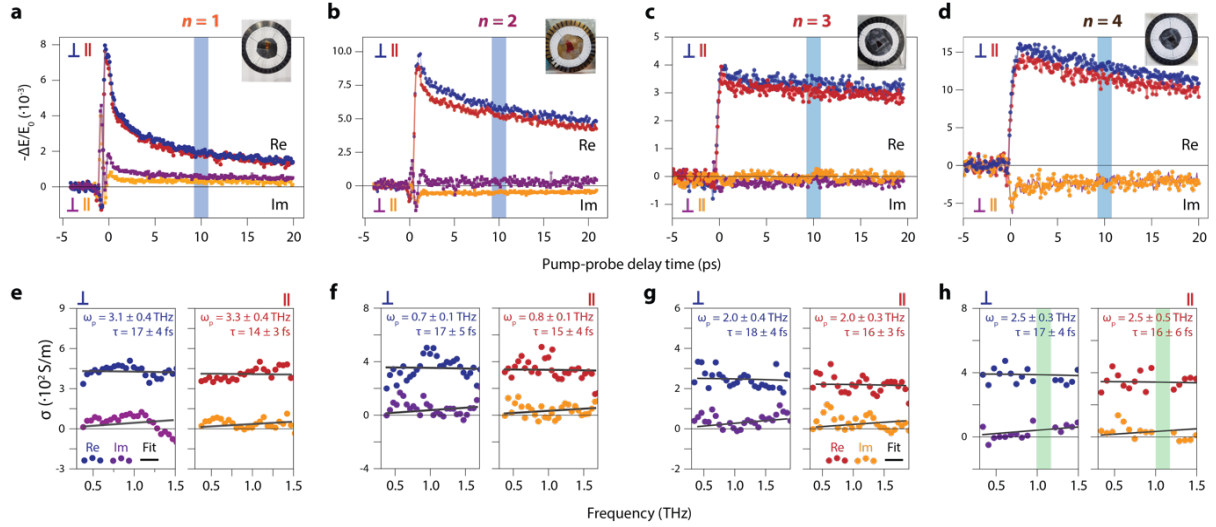

**Figure 4: Polarization-resolved optical-pump/THz-probe (OPTP) experiments on the 2D BAPI single crystals.** All samples are excited at 400 nm at low fluence (i.e. in the linear regime) and additional care was taken for each $n$ to measure the parallel and perpendicular photoconductivity at the same pump fluence. **(a)** Real and imaginary OPTP traces for $n = 1$. **(b)** Real and imaginary OPTP traces for $n = 2$. **(c)** Real and imaginary OPTP traces for $n = 3$. **(d)** Real and imaginary OPTP traces for $n = 4$. The blue vertical bars in **(a-d)** indicate at which pump-probe delay time we acquired conductivity spectra and averaged the OPTP signal. **(e-h)** Conductivity spectra for different $n$ measured at a pump-probe delay time of 10 ps. Left and right panels show conductivity spectra obtained with the THz polarization perpendicular and parallel to the 1 THz transition dipole moments, respectively. Solid lines are fits to the Drude model. In panel **(h)**, the vertical green bars indicate the spectral range in which the transmittance of the THz field was lower than 2%, which was omitted from further analyses.

The conductivity spectra along both directions for each thickness, $n$, are shown in Figure 4(e-h). Surprisingly, for all samples, we obtain a conductivity spectrum that does not resemble an excitonic response in our frequency range (a negative imaginary- and zero real photoconductivity, i.e., a Lorentzian oscillator for an interexcitonic transition) at 10 ps pump-probe delay time. Instead, the positive real, and almost zero imaginary photoconductivity for all $n$ looks Drude-like. Indeed, we are able to fit our data with the Drude model, from which we obtain the plasma frequency (proportional to the carrier density) and the carrier scattering time.

The results of all the photoconductivity experiments are summarized in Figure 5. We fitted the OPTP transients with a model that contains the sum of two decaying exponentials convolved with the IRF (except for $n = 4$, where we only needed one single decaying exponential to fit our data), of which the fitted decay rates are shown in Figure 5(a). All the fitted decay rates show little to no dependence on the direction in which we probe the photoconductivity. The slower decay rate, resulting from carrier recombination, goes down from $4.40 \pm 0.01 \cdot 10^{-2}$ ps$^{-1}$ for $n = 1$, to $6.32 \pm 0.06 \cdot 10^{-3}$ ps$^{-1}$ for $n = 3$, and increases again for $n = 4$ up to $1.8 \cdot 10^{-2}$ ps$^{-1}$.

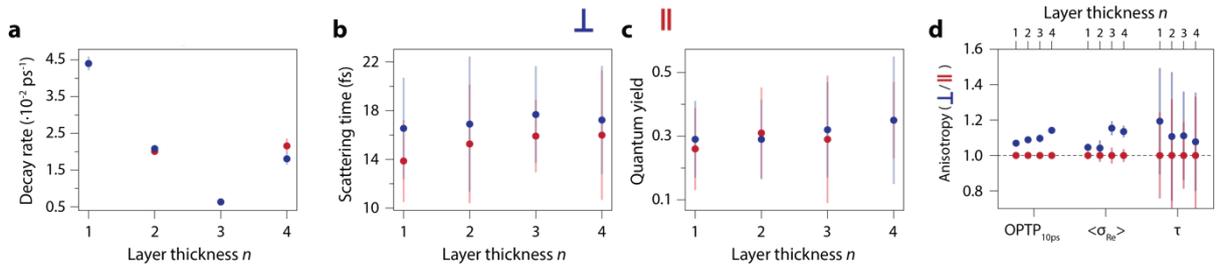

**Figure 5: Results from fitting the Drude model to the conductivity spectra and hints of anisotropy in the photoconductivity in 2D single crystalline BAPI.** In all panels the blue datapoints are data obtained with a THz polarization perpendicular to the 1 THz transition dipole moments, the red datapoints are obtained parallel to the 1 THz TDM. **(a)** Slowest decay rate vs. layer thickness, obtained from fitting the real part of the OPTP traces [Figure 3(a-d)]. **(b)** Scattering times vs. $n$. **(c)** Quantum yield vs. $n$, which was obtained by comparing the absorbed photon density to the carrier density calculated from the plasma frequency. **(d)** Averaged real part of the OPTP signal [blue vertical bars in Figure 3(a-d)], averaged real part of the conductivity spectra, and scattering times at 10 ps pump-probe delay time vs. $n$. Note how all the blue datapoints are higher for all $n$ than the red datapoints, indicating that the photoconductivity in 2D BAPI is ~10% higher in the direction perpendicular to the 1 THz TDM compared to parallel to it.

The anisotropy of photoconductivity in all four BAPI thicknesses is highlighted in Figure 5(b). The scattering times vs. the inorganic layer thickness $n$ are obtained from fitting the Drude model to the conductivity spectra [shown in Figure 4(e-h)]. The scattering times for all the different layer thicknesses are quite similar, around 17 fs, but we do, however, observe that in all our samples, the scattering time in the direction perpendicular to the 1 THz transition dipole moment (i.e. the <1 0 2> direction) is consistently higher than in the direction parallel to it (i.e. the <-2 0 1> direction), although they lie within each other's standard deviation from the fit; an indication that the photoconductivity is higher in one direction compared to the other. We also compare the photogenerated carrier quantum yields, defined as the carrier density obtained from the plasma frequencies, divided by photogenerated carrier density (obtained from the absorbed photon fluence), shown in Figure 5(c). Both the plasma frequencies and the quantum yields do not show a clear anisotropic trend vs. $n$, indicating that in both directions, the generated carrier densities are identical and excluding this as a cause for the apparent anisotropy in photoconductivity. The value of the quantum yield, around 30%, is in line with earlier observations in 3D perovskites[5,66].

To quantify the anisotropy in the photoconductivity, we compare various measured parameters in Figure 5(d), where we show the average of the real photoconductivity from the OPTP traces around 10 ps [shown by the vertical blue bars in Figure 4(a-d)], the average of the real part of the photoconductivity spectrum [shown in Figure 4(e-f)] and the scattering times obtained from the Drude fits as a function of $n$ and for both perpendicular and parallel directions to the 1 THz TDM. Here, we defined the anisotropy ratio as the value perpendicular to the 1 THz TDM divided by the value parallel to the 1 THz TDM. As can be seen, for all different $n$, the apparent photoconductivity is higher in the direction perpendicular to the 1 THz transition dipole moment, compared to parallel to it. From this, we estimate this anisotropy of the photoconductivity between the direction perpendicular and parallel to the 1 THz TDM to be around 10% for all samples.

To explain these findings, we turn to the Feynman variational polaron solution of the extended Fröhlich polaron model Hamiltonian. The original Feynman theory explicitly includes multiple phonon modes[67], and anisotropic effective masses[68]. Here, we use the calculated anisotropic modes in the LT phase, associate a Fröhlich dielectric mediated electron-phonon coupling with the infrared activity of each mode, and solve the finite temperature mobility theory for 300 K.

Due to the varying infrared activity in the two in-plane directions, we find a Fröhlich dimensionless electron-phonon coupling of 3.44 and 3.75 along the two in-plane lattice vectors; both considerable higher than in 3D halide perovskites. This leads to predicted mobilities of 1.89 cm$^2$/V/s and 1.76 cm$^2$/V/s in the two directions, an anisotropy of 8%, a value in line with the experimentally obtained anisotropy estimates. The additional electron-phonon coupling also increases polaron localization, by a similar quantity. We, therefore, explain the experimentally observed anisotropy in photoconductivity to be due to different intrinsic carrier mobilities, arising directly from the anisotropy in the dielectrically mediated electron-phonon coupling strength. We cannot, however, disentangle the effects of scattering time and effective mass, on the carrier mobility. Overall, it is clear that the impact the 1 THz phonon mode has on the electronic degrees of freedom is remarkable[25].

CONCLUSIONS

We have studied anisotropy in the phonon transition dipole moments and the optical conductivity of single crystalline two-dimensional butylammonium lead iodide perovskites. From the linear THz absorption measurements, we determine the transition dipole moment vectors of the 1 THz modes to lie in the <1 0 2> / <2 0 1> family of directions, show that the 2 THz mode is angle independent and discuss similarities and differences between the vibrational modes with different thicknesses of the inorganic layer $n$. Furthermore, we measured the optical conductivity by optical-pump/THz probe spectroscopy, and show that the photoconductivity is higher in the direction perpendicular to the 1 THz transition dipole moment by 5-10%, compared to the direction parallel to it. Theoretical calculations based on Feynman's variational polaron theory corroborate the observed anisotropy and pinpoint that directional electron-phonon interactions are likely responsible for this effect. The impact of nuclear displacement along the 1 THz phonon coordinate on the electronic degrees of freedom in metal-halide perovskites is remarkable. Our results shed new light on some of the fundamental molecular physics governing direction-dependent effects in these quasi-two-dimensional perovskite semiconductors and corroborate the importance of the dynamic interplay between vibrational modes and charge carrier motion in these materials.

**Supporting Information available.** Detailed description of the interfacial synthesis method, description of experimental techniques, data analysis of the THz TDS and OPTP experiments, description of simulation methods, supplementary table S1-S4, and supplementary figures S1-S13.

**Competing financial interests.** The authors declare no competing financial interests.

**Corresponding authors.** Mischa Bonn: bonn@mpip-mainz.mpg.de

Heejae Kim: heejaekim@postech.ac.kr

Jaco J. Geuchies: j.j.geuchies@lic.leidenuniv.com

**Acknowledgements.** JJG gratefully acknowledges funding from the Alexander von Humboldt Stiftung. JMF is funded on Royal Society grant URF-R1-191292. LDV acknowledges the European Union's Horizon 2020 research and innovation program under the Marie Sklodowska-Curie grant No 811284 (UHMob). J. K. acknowledges support from the Swedish Research Council (VR) program 2021-00486. Computations were enabled by resources provided by the National Academic Infrastructure for Supercomputing in Sweden (NAISS) at NSC and PDC partially funded by the Swedish Research Council through grant agreement no. 2022-06725. We thank all members of the THz group of the molecular spectroscopy department at MPIP for many fruitful discussions.

TOC

THz spectroscopy & simulations reveal **anisotropy** in:

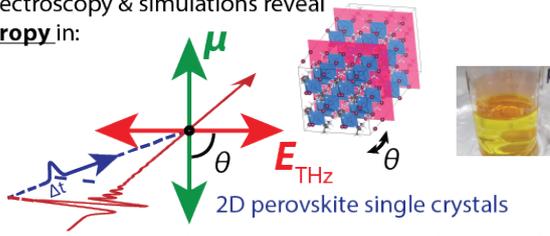

2D perovskite single crystals

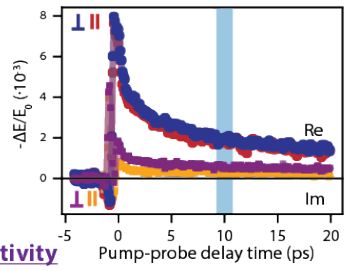

**Vibrational transition dipole moments**      **Photoconductivity**

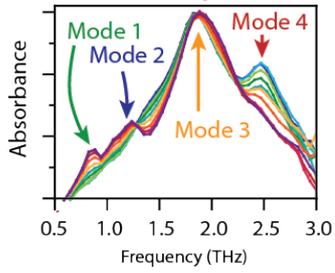

due to
**electron-phonon interactions**

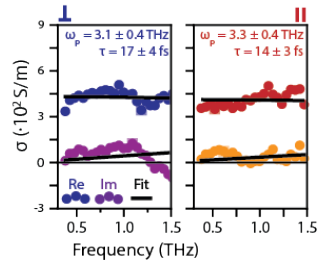

# Supporting information

# Anisotropic electron-phonon interactions in 2D lead-halide perovskites

Jaco J. Geuchies[1*], Johan Klarbring[2,3], Lucia Di Virgillio[1], Shuai Fu[1], Sheng Qu[1], Guangyu Liu[4], Hai Wang[1], Jarvist M. Frost[4], Aron Walsh[2], Mischa Bonn[1*] and Heejae Kim[1,5*]

1. Max Planck Institute for Polymer Research, 55128 Mainz, Germany.

2. Department of Materials, Imperial College London, London SW7 2AZ, United Kingdom.

3. Department of Physics, Chemistry and Biology (IFM), Linköping University, SE-581 83, Linköping, Sweden

4. Department of Physics, Imperial College London, London SW7 2AZ, United Kingdom.

5. Department of Physics, Pohang University of Science and Technology, 37673, Pohang, Korea.

*Corresponding authors

# Table of Contents

## SI1 - Materials and methods

**Chemicals** PbO (>99.9%), hydroiodic acid (57% w/w in water), hypophosphorous acid ($H_3PO_2$, 50% w/w in water), n-butylamine (99.5%) were purchased from Sigma-Aldrich. CH3NH3I was obtained fromLuminescence Technology Corp. Methylammonium chloride (low water content) was bought from TCI. All chemicals were used without further purification.

**Synthesis procedure.** The synthesis of cm-sized crystal flakes was done at a liquid-air interface, based on an adjusted protocol by Wang *et al.*[1]. For the exact amounts, see table S1 below. We mixed PbO in HI (57% w/w solution in water) and $H_3PO_2$ (50% w/w solution in water) in a 20mL glass vial. The mixture was heated to 80°C, at which point the PbO dissolved and a clear yellow solution was obtained. The thermocouple that controls the hot-plate temperature was always inserted in a 20 mL glass vial filled with the same volume of water as the perovskite precursor solutions.

Next to this, a solution of n-butylamine (nBAm) and methylammonium (MA) chloride in HI was prepared while cooling the vial, holding the nBAm and MA, in an icebath. Care should be taken to add the HI slowly, as the reaction is very exothermic (one can observe fumes forming as the HI is added). This nBAm/MA solution is dropwise added to the Pb-containing solution. Upon mixing the two solutions, some small crystallites are formed, which dissolve rapidly.

Depending on the layer thickness that is formed, crystal growth is done either by lowering the temperature from 80°C to 70°C for *n*=1 and *n*=2, and crystal growth takes 10-30 minutes. For *n*=3 and *n*=4, we used a temperature controller (JKEM, controller type T), with a glass-coated probe inserted in a vial filled with water on the same hot plate, and ramped the temperature down by 2°C per hour until nucleation of a crystal at the liquid-air interface, at which point the temperature was kept constant (usually between 55-65°C).

After the growth of the crystals at the liquid-air interface, the crystal is scooped gently off the interface with Teflon tweezers, placed on a plastic lid, and dried in a vacuum oven for 5 hours. It is important to attach a liquid N2 cold-trap in between the oven and the vacuum pump, to catch off the remaining water-HI solution. After drying, the samples are stored in a nitrogen-filled glovebox until further use. Note that the overall crystal thickness varies from 10-20 micrometers for *n*=1 (making them very fragile) up to 200 micrometers for *n*=4.

Additionally, for the layers where n > 1, we used MACl (instead of MAI). For reasons unclear to us, we were able to get interfacial nucleation more consistently (instead of nucleation in the bulk of the liquid), using MACl compared to MAI.

Precursor quantities for the synthesis of $(BA)_2(MA)_{n-1}Pb_nI_{3n+1}$ single crystals

**Table S1:** amount of precursors used in the synthesis of the $(BA)_2PbI_4$ crystals (n=1)

| n=1 | Chemical | Amount (mmol) | Amount (mg) | Amount (mL) |
|---|---|---|---|---|
| PbI solution | PbO powder | 5 | 1116 | |
| | 57% w/w acqueous HI solution | 38 | | 5 |
| | 50% (w/w I think) acqueous $H_3PO_2$ solution | 7.75 | | 0.85 |
| n-$CH_3(CH_2)_3NH_3I$ solution | n-$CH_3(CH_2)_3NH_2$ (liquid) | 5 | | 0.462 |
| | 57% w/w acqueous HI solution | 19 | | 2.5 |

**Table S2:** amount of precursors used in the synthesis of the $(BA)_2(MA)Pb_2I_7$ crystals (n=2)

| n=2 | Chemical | Amount (mmol) | Amount (mg) | Amount (mL) |
|---|---|---|---|---|
| PbI solution | PbO powder | 5 | 1116 | |
| | 57% w/w acqueous HI solution | 38 | | 5 |
| | 50% (w/w I think) acqueous $H_3PO_2$ solution | 7.75 | | 0.85 |
| n-$CH_3(CH_2)_3NH_3I$ solution | $CH_3NH_3Cl$ | 2.5 | 169 | |
| | n-$CH_3(CH_2)_3NH_2$ (liquid) | 7 | | 0.694 |
| | 57% w/w acqueous HI solution | 19 | | 2.5 |

**Table S3:** amount of precursors used in the synthesis of the $(BA)_2(MA)_2Pb_3I_{10}$ crystals (n=3)

| n=3 | Chemical | Amount (mmol) | Amount (mg) | Amount (mL) |
|---|---|---|---|---|
| PbI solution | PbO powder | 5 | 1116 | |
| | 57% w/w acqueous HI solution | 38 | | 5 |
| | 50% (w/w I think) acqueous $H_3PO_2$ solution | 7.75 | | 0.85 |
| n-$CH_3(CH_2)_3NH_3I$ solution | $CH_3NH_3Cl$ | 3.33 | 225 | |
| | n-$CH_3(CH_2)_3NH_2$ (liquid) | 1.67 | | 0.164 |
| | 57% w/w acqueous HI solution | 19 | | 2.5 |

**Table S4:** amount of precursors used in the synthesis of the $(BA)_2(MA)_3Pb_4I_{13}$ crystals (n=4)

| n=4 | Chemical | Amount (mmol) | Amount (mg) | Amount (mL) |
|---|---|---|---|---|
| PbI solution | PbO powder | 5 | 1116 | |
| | 57% w/w acqueous HI solution | 38 | | 5 |
| | 50% (w/w I think) acqueous $H_3PO_2$ solution | 7.75 | | 0.85 |
| n-$CH_3(CH_2)_3NH_3I$ solution | $CH_3NH_3Cl$ | 3.75 | 253.5 | |
| | n-$CH_3(CH_2)_3NH_2$ (liquid) | 1.25 | | 0.124 |
| | 57% w/w acqueous HI solution | 19 | | 2.5 |

## SI2 - Additional steady-state characterization

X-ray diffraction was performed on a Rigaku Smartlab diffractometer, using copper Kα radiation (1.54 angstrom), both in Bragg-Brentano geometry for powder-XRD and in transmission geometry. From the powder XRD patterns, we fitted the peak position of the reflections. Since we are only sensitive to specular reflections in the out-of-plane directions in this geometry (i.e. reflections from the inorganic lamella), we could determine the interlayer spacing for each of the different thicknesses of BAPI.

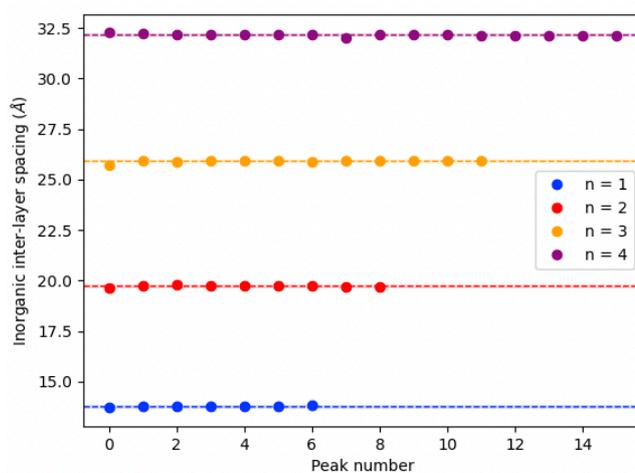

**Figure S1: from the fitted p-XRD peak positions, we calculate the interlayer spacing, as shown and disussed in the main text.**

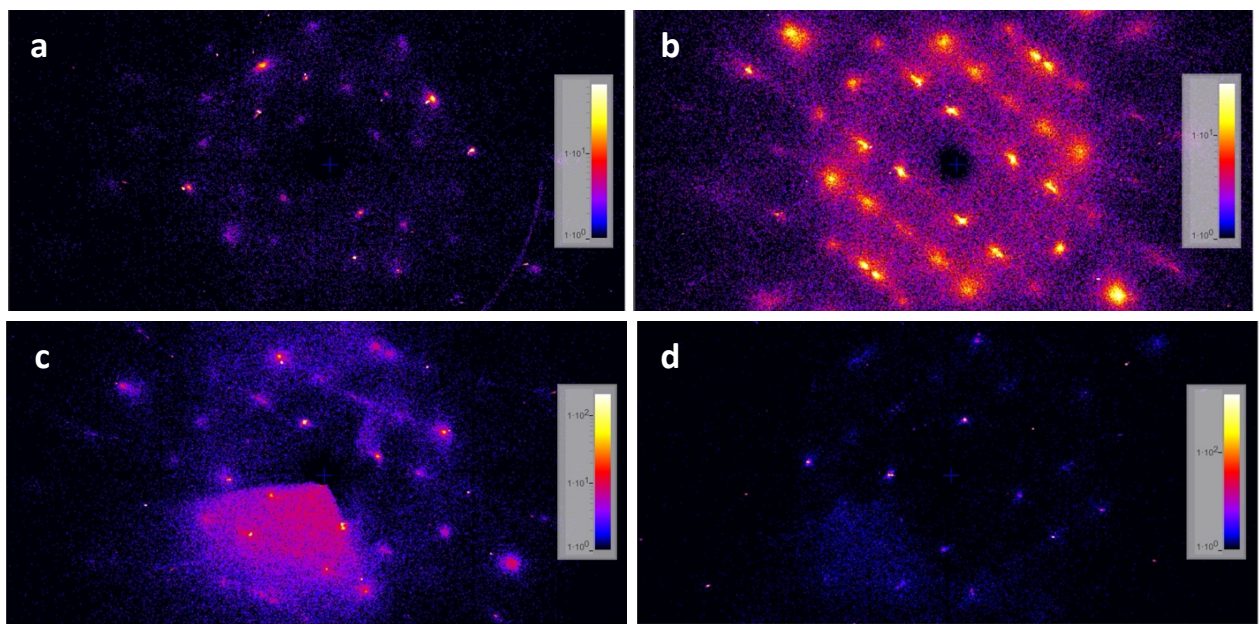

**Figure S2: transmission 2D WAXS patterns from the samples shown throughout the main text. (a)** n = 1 BAPI. **(b)** n = 2 BAPI. **(c)** n = 3 BAPI. **(d)** n = 4 BAPI. The triangular shaped wedge in panel **(c)** originated from grease that was unfortunately present on the Kapton foil that holds the beamblock for the direct beam.

Optical reflectivity curves were measured with a Perkin-Elmer Lambda 900 UV/Vis/NIR spectrometer, where samples were placed at the direct end of an integrating sphere to collect all reflected light.

The thickness of the samples was measured using a Dektak profilometer.

## SI3 - Optical-pump/THz-probe spectroscopy

Here, we describe the technique THz time-domain spectroscopy (TDS) and Optical Pump THz Probe Spectroscopy (OPTP). This technique is able to probe the conductivity of carriers and further retrieve the charge carrier mobility $\mu$.

We use an amplified Ti:sapphire laser producing pulses with 800 nm central wavelength and ~50 fs pulse duration at 1 kHz repetition rate. The THz field is generated by optical rectification in a ZnTe(110) crystal (thickness 1 mm). The THz detection is based on the electro-optic (also called Pockels) effect in a second ZnTe crystal with 1 mm thickness. We vary the time delay between the THz field and the 800 nm sampling beam with a motorized delay stage (M-605.2DD purchased from Physik Instrument (PI)). The time delay between the optical pump and THz probe pulses is controlled by a second motorized delay stage (M521.DD, Physik Instrument (PI)). The pump pulse has a 400 nm central wavelength and is produced by the second harmonic generation of the 800 nm femtosecond pulse in a beta barium borate (BBO) crystal, where the remaining 800nm is filtered afterwards out with a short-pass filter (blue colorglass). For the excitation-wavelength dependent measurements, we used an optical parametric amplifier (OPA) to convert our 800nm fundamental beam into the various pump wavelengths, resonant with the band-edge excitonic transition of our materials, which we filter further using a bandpass filter for the corresponding wavelength directly after the OPA.

**Sample stability during pump-probe measurements:** For the OPTP measurements, we verify we are in the low-fluence or linear regime. Furthermore, at higher fluences we observe sample degradation during the measurement, over the course of roughly two hours, which can be observed by eye by the loss of photoluminescence from the sample. After checking initial samples to find a suitable stability window in terms of fluence, the data presented throughout this paper is optimized such that we do not have loss of photoluminescence, which seemed to be a good indicator for sample stability. Furthermore, we do note that spincoated films of BAPI photodegrade much faster than the single crystals we show here.

## SI4 - Analytical model for the transmittance of THz pulses and photoconductivity through the quasi-2D perovskite single crystals

We do *not* use the Tinkham approximation[2], i.e. the thin-film approximation, to calculate the photoconductivity spectra. The reason for this is that 1) the crystals are relatively thick (tens to hundreds of micrometers), and 2) the ground-state response in the THz frequency range is not flat due to the presence of vibrational resonances[3–6]. Instead, we write down the analytical Fresnel equations for the complex transmission of a THz field at a given frequency:

$$T_{calc}(f) = \frac{E_{sample}}{E_{free}} = \frac{t_{01} p_1 t_{10} FP_{010}}{p_0} \quad \text{(eq. S1)}$$

With $T_{calc}$ the complex transmission at a given frequency $f$, $E_{sample/free}$ the amplitude of the THz electric field over a length $L$ (the thickness of the crystal, measured with a profilometer) with the free-standing single crystal sample or without anything in its' path respectively. The transmission through the air-perovskite interface, $t_{01}$, propagation through the perovskite crystal, $p_1$, transmission through the perovskite-air interface, $t_{10}$, and propagation of the THz pulse through the sample or air over a length L are specified. Furthermore, due to the high refractive index of the perovskite material, Fabry-Perot-like multiple reflections are possible, captured by the term $FP_{010}$. These Fresnel coefficients are given by:

$$t_{01} = \frac{2}{1+n_c} \qquad t_{10} = \frac{2n_c}{1+n_c} \qquad P_1 = e^{\frac{i2\pi f L n_c}{c}} \qquad P_0 = e^{\frac{i2\pi f L}{c}}$$

$$r_{10} = r_{01} = \frac{n_c - 1}{n_c + 1} \qquad FP_{010} = \frac{1}{1 - r_{10}^2 p_1^2}$$

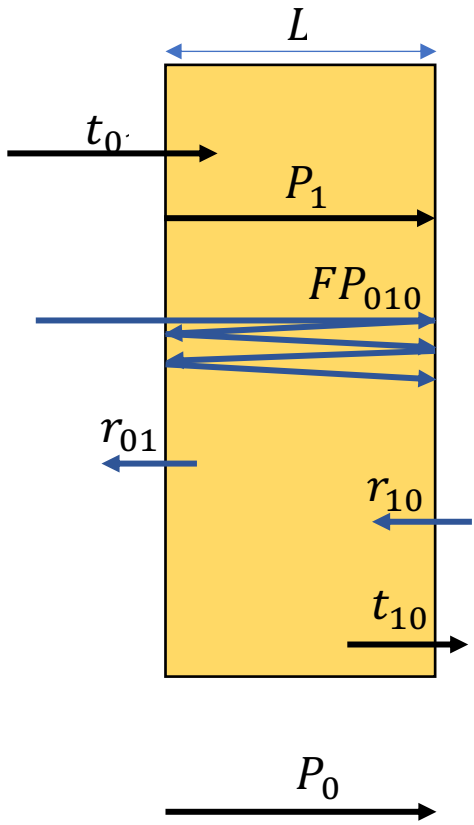

**Figure S3: schematic of the transmission pathways of a photon at a given frequency through a sample of thickness $L$.**

Here, in the Fabry-Perot term, we have taken the limit of an infinite number of internal reflections through the sample, which converges the geometric sum. We believe this is a valid approximation as each increasing term in this sum gets significantly smaller due to absorption through the material. We use equation S1 to numerically retrieve the refractive index of our perovskite single crystal in the THz frequency range, for all the different sample orientations we measured, in both the ground- and photoexcited state.

# SI5 - Numerical minimization to obtain complex refractive indices and photoconductivity in the ground- and photoexcited state

**Data pre-treatment: acquiring the proper phase.** We now use the experimentally obtained complex transmission to obtain the real and imaginary parts of the complex refractive index. Since we measure the electric field of our THz pulse, we directly are able to measure both amplitude and phase, as shown below.

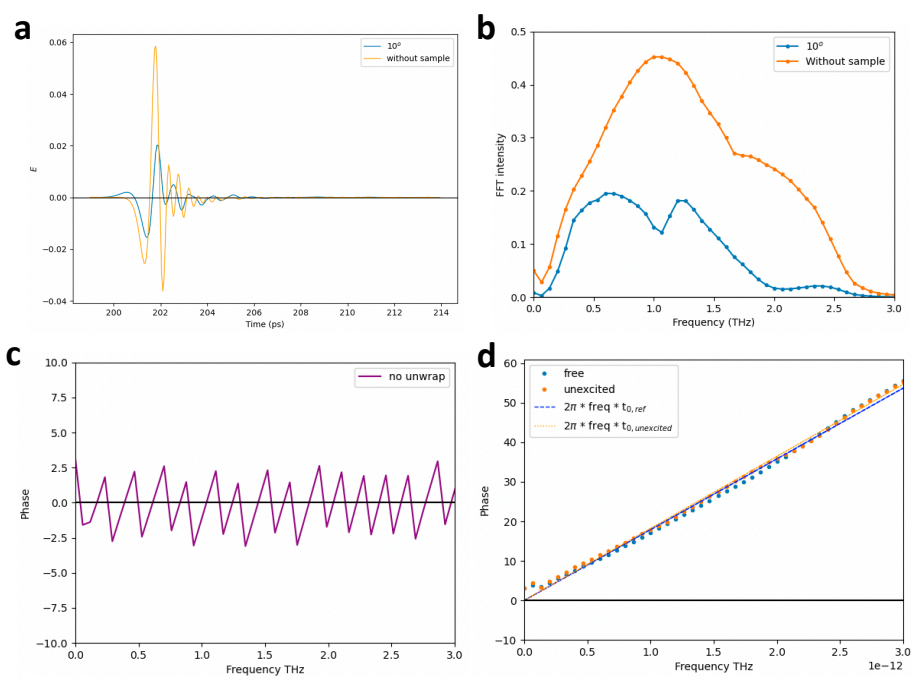

**Figure S4: THz-TDS data and data analysis for n = 1. (a)** Example of measured THz waveforms wit hand without the single-crystal in the THz beampath. **(b)** Absolute value FFT of the data shown in **(a)**, giving us the amplitude spectrum. Due to the relative thickness of the film, the vibrational modes are already clearly observed. **(c)** Raw, (incorrectly) unwrapped, phase of the THz fields, defined as the arctan(Imaginary/Real) for each frequency of the Fourier-transformed data from **(a)**. **(d)** Properly unwrapped phase as a function of frequency.

We obtain the experimental complex transmission as a function of frequency of our data, shown in Figure S4 above, as $T_{sample}$ = FFT($E_{sample}$)/FFT($E_{free}$). We can now rewrite our complex transmission with the properly unwrapped phase, which is important for the numerical minimization further ahead:

$$T_{sample} = |T_{sample}|e^{i\phi} \quad \text{(eq. S2)}$$

with $\phi$ being the unwrapped phase.

**Numerical minimization.** Since we now have the complex transmission of the sample, we can numerically minimize the mismatch of the calculated value (from equation S1) against our experimental data. To this end, we compute the difference between the calculated and measured amplitudes and phases, $\Delta T$ and $\Delta \phi$ respectively, and sum up the squared difference between experiment and calculation over all frequencies in our THz window:

$$difference = \sum_{frequencies} \Delta T^2 + \Delta \phi^2 \quad \text{(eq. S3)}$$

This difference can be numerically minimized against the complex refractive index of the sample $n_c$ through various strategies. We opted to use a double minimization scheme, starting with a Basinhopping minimizer, implemented in the scipy.optimize library of Python. This minimizer starts at different initial guesses for the refractive index, for which we used the refractive index from the Tinkham approximation as an initial guess, and let the minimizer run for either a fixed or predetermined number of minimization steps, an example of which is shown in Figure S5 below. The advantage of the Basinhopping method, is that it searches a large parameter space for a global minimum.

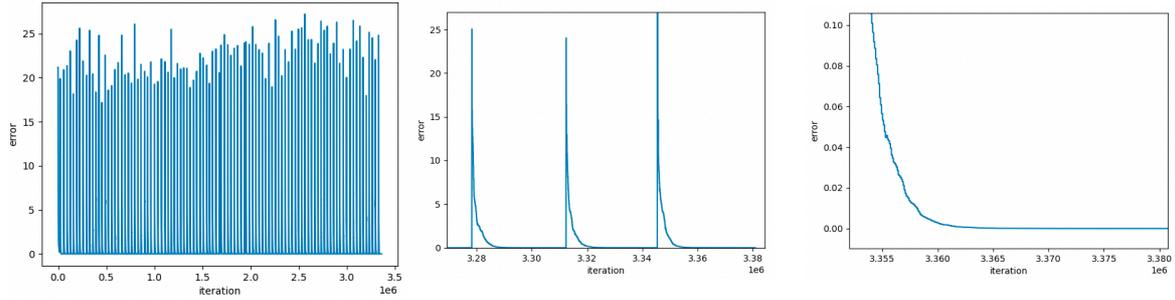

Figure S5: example of Basinhopping minimization, where from left to right, we zoom in more into the minimization steps.

The output of the Basinhopping algorithm does not check for convergence, so we run the minimized complex refractive index obtained through Basinhopping through a similar minimizer, scipy.minimize, using a Nelder-Mead algorithm to check for convergence.

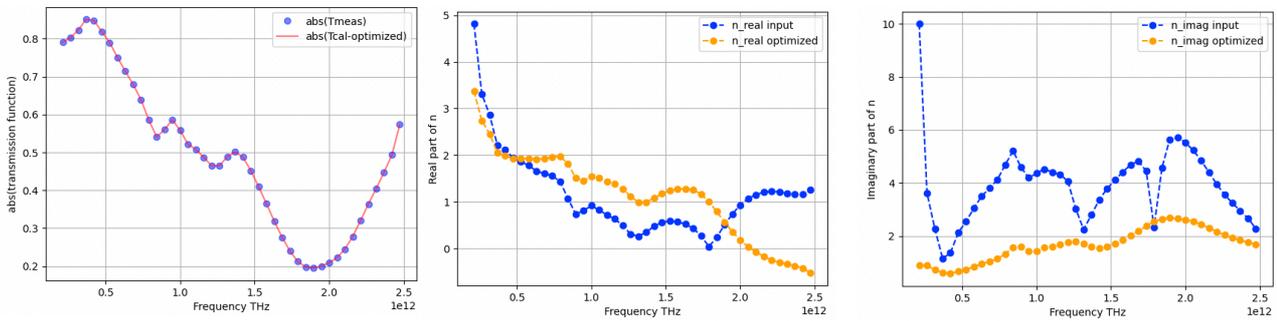

**Figure S6: Example of minimized refractive index.** The left panel shows the absolute transmission for both the experimental data, and through calculation with the minimized refractive index of the sample. The middle and right panels show the minimized real and imaginary parts of the refractive indices respectively, with in blue the initial guess and in yellow the minimized refractive index.

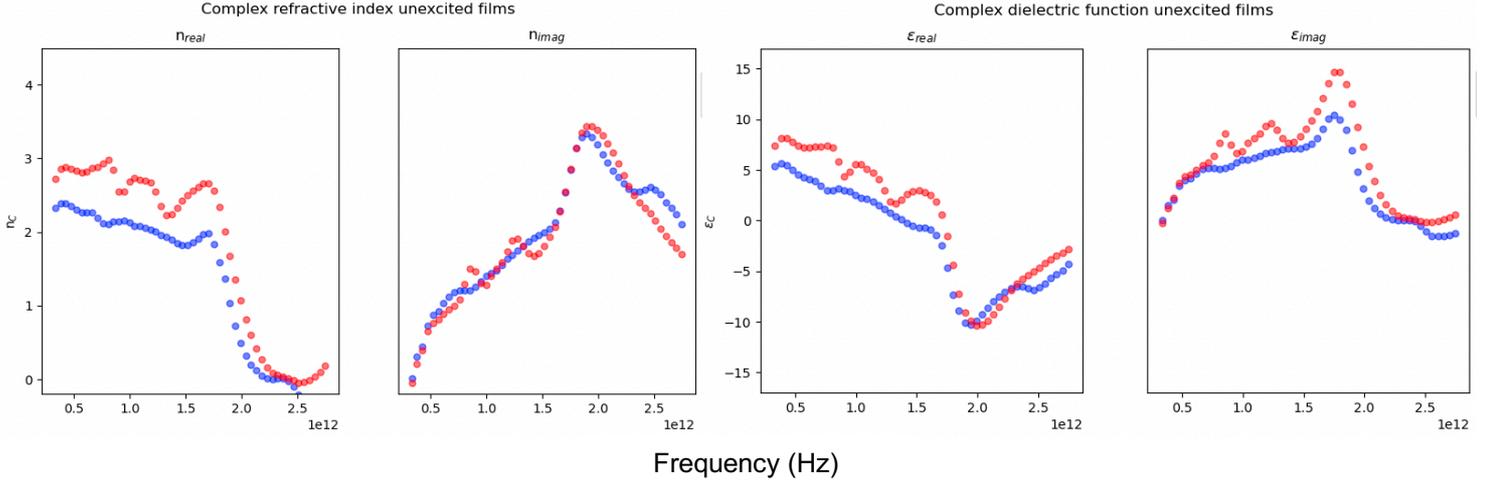

**Figure S7: Imaginary and real parts of the refractive index (left) for the two extreme angles (90º apart) for n = 1 BAPI, which can be converted into complex dielectric functions (right).**

We can apply the same minimization scheme to obtain the refractive index of the photoexcited sample. For this we calculate the electric field of the THz pulse transmitted through the photoexcited sample as

$$E_{photoexcited}(t) = E_{ground\ state}(t) + \Delta E(t) \quad \text{(eq. S4)}$$

Where we add the measured photoinduced change of the THz field, $\Delta E(t)$, to the transmitted field without photoexcitation $E_{ground\ state}(t)$. For the minimization, as an initial guess for the to-be optimized refractive index, we use the ground-state refractive index as an initial guess.

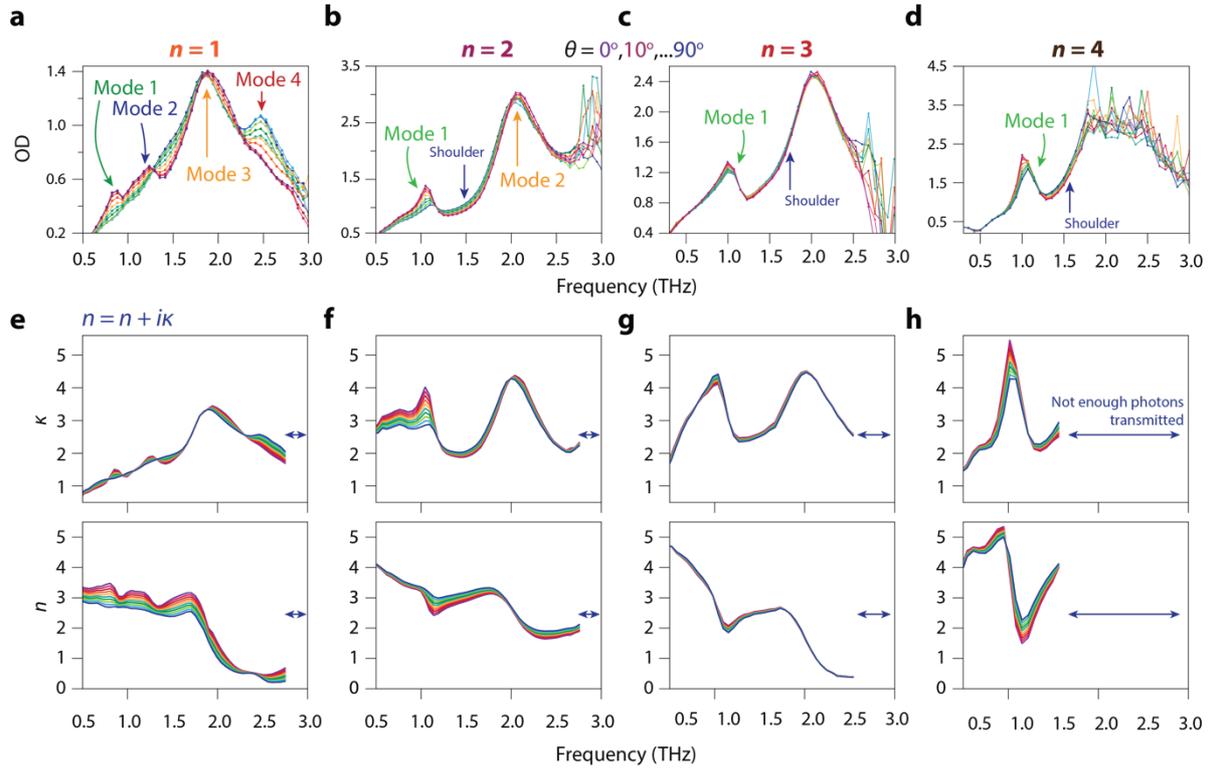

**Figure S8: retrieved complex refractive indices in the THz spectral range for n = 1 – 4 BAPI. (a-d)** optical density spectra for n = 1 to n = 4 going from left to right, as a function of angle. **(e-h)** Complex refractive indices for n = 1 to n = 4 BAPI going from left to right, as a function of angle. The top panels show the imaginary part of the refractive index, the bottom panels show the real part. The horizontal blue arrow indicates the spectral region that has less than 5% transmission of the THz field, which we omit from the numerical minimization due to the reduced signal-to-noise.

**Obtaining the photoconductivity spectra.** We convert the numerically retrieved refractive indices into dielectric functions, which are additive, and separate them into various terms.

$$n_c^2 = \varepsilon_c$$
$$\varepsilon_c = \varepsilon_{real} + i\varepsilon_{imag}$$

$$\varepsilon_{real} = n^2 - \kappa^2 \qquad \text{(eq. S5)}$$

$$\varepsilon_{imag} = 2n\kappa \quad \text{(eq. S6)}$$

Once we have the complex dielectric function of the sample $\varepsilon_c$, we decompose the total dielectric response of our system in the ground and photoexcited state as

$$\varepsilon_{c,groundstate} = \varepsilon_\infty + \varepsilon_{lattice} \quad \text{(eq. S7)}$$
$$\varepsilon_{c,excited\ state} = \varepsilon_\infty + \varepsilon_{lattice} + \varepsilon_{carriers} \quad \text{(eq. S8)}$$

Where we separate the total dielectric function into a constant contribution $\varepsilon_\infty$, a lattice response $\varepsilon_{lattice}$ (i.e. vibrational modes that are present), and in the photoexcited state an additional response from the generated carriers $\varepsilon_{carriers}$. The latter term we want to isolate to obtain the conductivity of the photoexcited carriers: $\varepsilon_{carriers} = \varepsilon_{c,excited\ state} - \varepsilon_{c,groundstate}$. We convert this into a photoconductivity spectrum

$$\sigma(\omega) = -i\omega\varepsilon_0\varepsilon_{carriers}(\omega) \quad \text{(eq. S9)}$$

The photoconductivity spectra that we retrieved via this method are shown in Figure 4 of the main text and are fitted with the Drude model:

$$\sigma(\omega) = \frac{\omega_p^2 \varepsilon_0 \tau_s}{1 - i\omega\tau_s} \quad \text{(eq. S10)}$$

With $\omega_p^2$ the squared plasma frequency (proportional to the carrier density) and $\tau_s$ the Drude scattering time. This method has the advantage that the acquired photoconductivity spectra are analytical, and do not rely on various assumptions, e.g. the Tinkham/thin-film approximation, for the transmitted THz pulse.

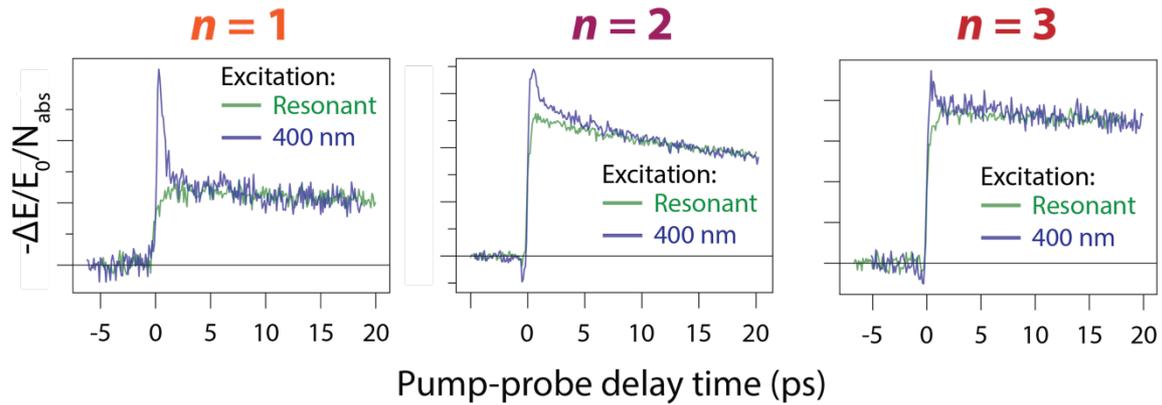

**Figure S9: Hot carriers are more mobile than cold carriers – origin of the initial fast decay.** OPTP traces for $n = 1$ (left), $n = 2$ (middle) and $n = 3$ (right) upon resonant band-edge excitation (green) and 400 nm excitation. Resonant excitation shows an IRF limited ingrowth of the OPTP transients, whereas 400 nm above-bandgap excitation shows a fast initial decay. Since the data is obtained at low photon fluences, i.e. in the linear regime, and normalized on the absorbed photon fluence, we can conclude that the initial decay originates from a higher mobility for hot carriers.

## SI6 - Computational details

The density functional theory (DFT) calculations were performed using VASP[7–9] and the projector augmented wave (PAW)[10] methodology, with the PBE[11] exchange-correlation functional with added Tkatchenko-Scheffler[12] dispersion corrections (PBE+TS). For structural relaxation and phonon calculations, we used an 800 eV cutoff energy, a 6x6x2 Γ-centered k-point grid (where the simulation cell was aligned so that the c-direction corresponds to the out-of-plane direction), a convergence criterion for the electronic self-consistent iterations of $10^{-8}$ eV and a Gaussian smearing of the electronic states of 10 meV. Full structural relaxation was performed until the force on any atom was less than 0.5 meV/Å. For each element we used the VASP-recommended PBE PAW-potentials, labeled 'Pb_d', 'I', 'N', 'H', 'C', and evaluated projectors in reciprocal space (LREAL = .FALSE.). The starting points for the structural relaxations of the n=1, LT and HT phases were taken from Ref. [13].

Harmonic phonon dispersion relations of the n=1, LT and HT phases were computed using the small-displacement formalism as implemented in Phonopy[14,15]. We used 2x2x1 supercells and a small-displacement distance of 0.01 Å.

Born effective charges (BECs) and dielectric tensors were calculated using VASPs density functional perturbation theory (DFPT) routines (LEPSILON = .TRUE.). Using the calculated BECs and Γ-point phonon frequencies and eigenvectors, IR spectra were obtained using the phonopy-spectroscopy package[16]; https://github.com/skelton-group/Phonopy-Spectroscopy]. We have cross-checked the small displacement phonon calculations for the LT phase with Γ-point eigenvectors and frequencies obtained directly from the VASP DFPT routines (LEPSILON = .TRUE. , IBRION=8), and the resulting IR-spectra are fully consistent.

For the electronic band structure and effective mass calculations we have included spin-orbit coupling (SOC) effects, used a 600 eV cutoff, a $10^{-6}$ eV electronic convergence criterion, and evaluated projectors in real space (LREAL = Auto). All other parameters were the same as above. Plotting of the electronic bands and extraction of effective masses was done using the sumo package[17].

## Phonon band structures and IR spectrum

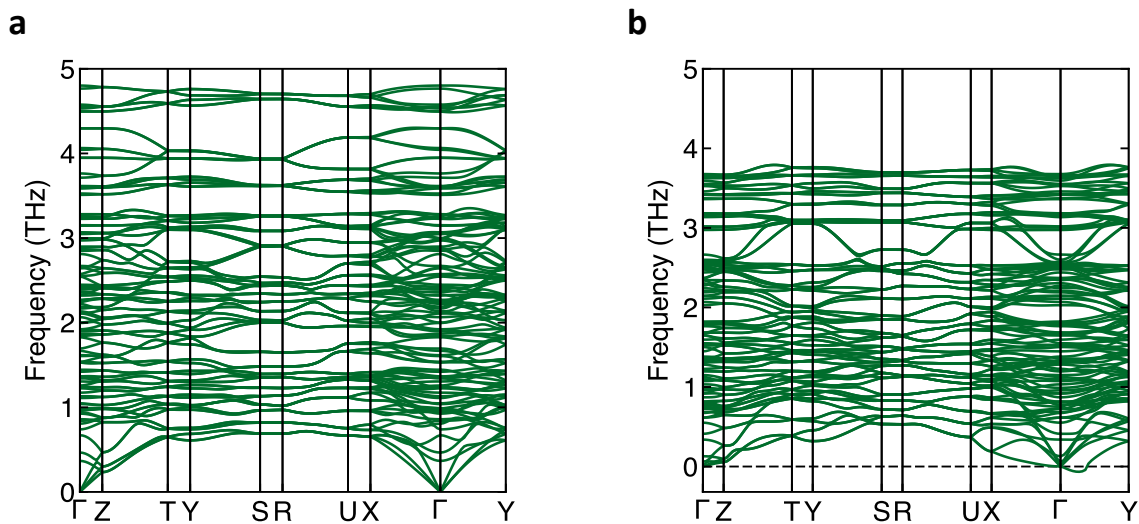

**Figure S10:** Low-frequency parts of the phonon dispersion relations of the (a) LT and (b) HT phases. The small imaginary mode near the Γ-point in the HT phase is likely a numerical artifact. In both phases, the Γ-Z corresponds to the out-of-plane direction while the Γ-X and Γ-Y corresponds to directions along the shorter and longer in-plane lattice vectors, respectively.

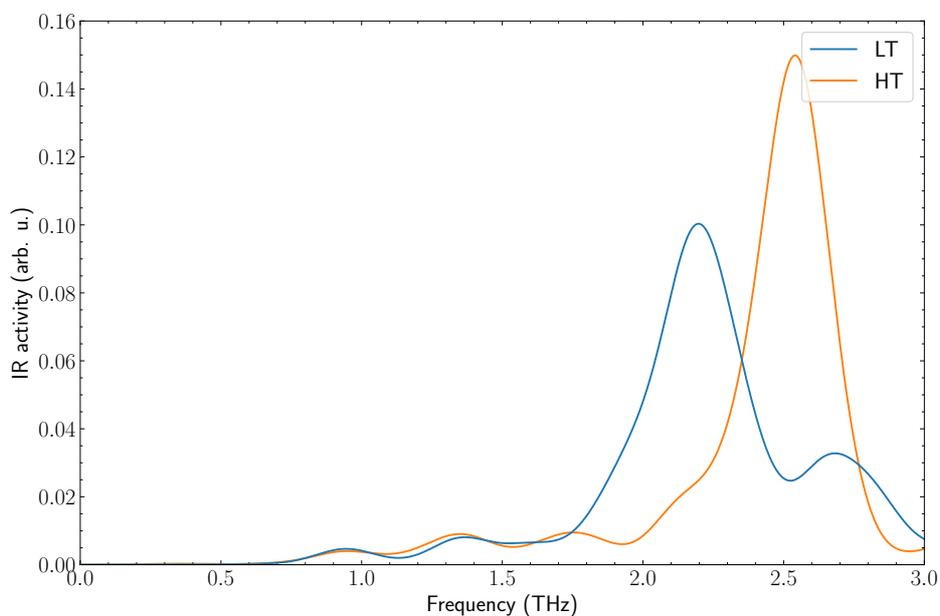

**Figure S11:** Comparison of the low frequency part of the calculated IR-spectrum for the n=1 LT and HT phases.

## Directionally resolved IR spectrum

We calculated a directionally resolved IR-spectrum by projecting the mode-effective charges on a unit-vector initially along the direction of the (a) shortest lattice vector, summing over all modes, and then letting this unit-vector rotate in-plane.

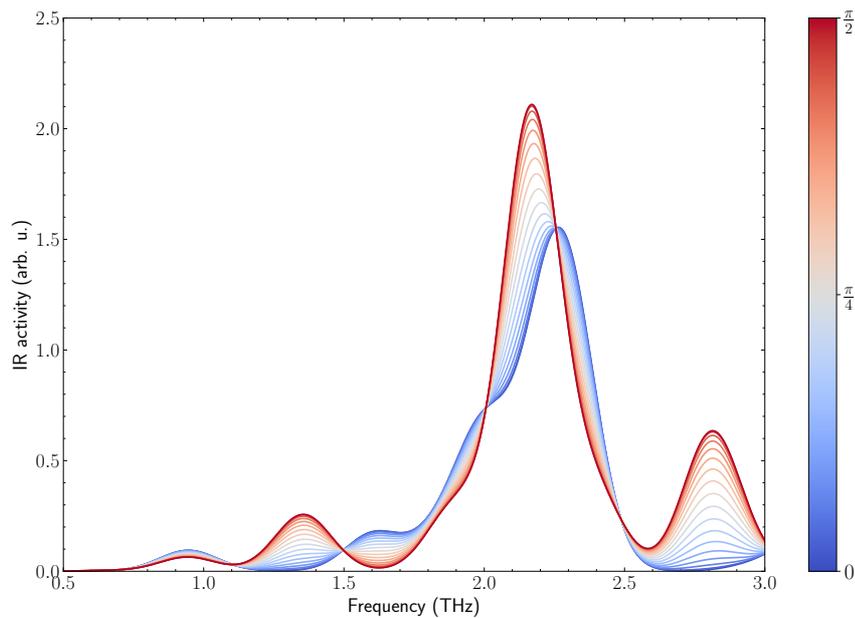

**Figure S12:** Directionally resolved IR-spectrum. The color scale gives the direction of the external field, in terms of positive in-plane rotation away from the shortest (a) in-plane lattice vector.

Electronic band structure

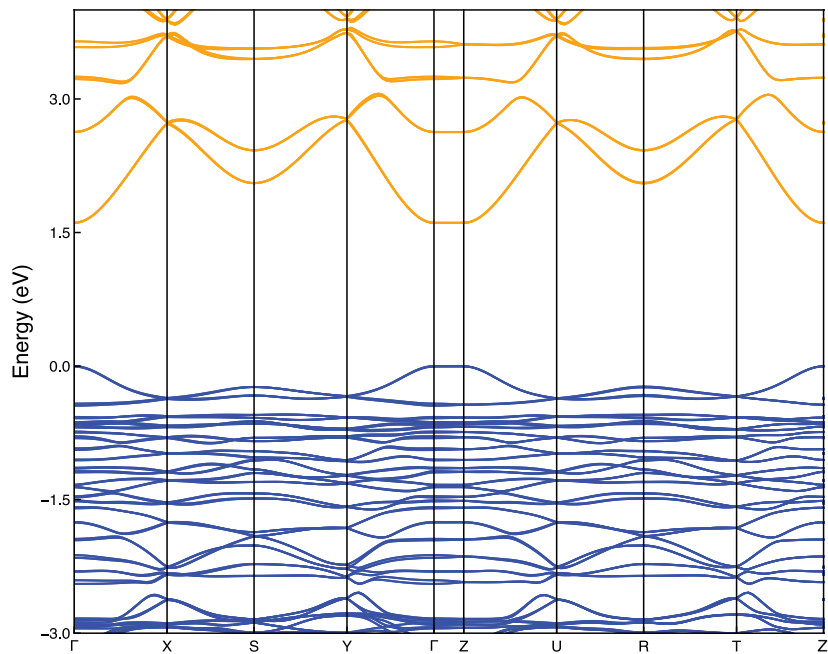

**Figure S13:** Electronic bandstructure, calculated using PBE+TS including SOC, of the n=1 LT phase. Parabolic fits to the band edges yields the following effective masses (in bare electron masses). Electrons: Γ-X: 0.249 , Γ-Y: 0.226. Holes: Γ-X: 0.396, Γ-Y: 0.397. Γ-Z corresponds to the out-of-plane direction while Γ-X and Γ-Y corresponds to directions along the shorter and longer in-plane lattice vectors, respectively.

## SI7 - Supplementary references